\documentclass[aps,pra,twocolumn,showpacs,amsmath,amssymb,preprintnumbers,superscriptaddress,10pt,longbibliography]{revtex4-1}

\usepackage{graphicx}
	\graphicspath{{figures/}}
\usepackage[binary]{SIunits}
\usepackage[bookmarks=false]{hyperref}
\usepackage{color}
\usepackage[capitalise]{cleveref}
	\crefname{section}{Sec.}{Secs.}% APS style uses abbreviations
	\Crefname{section}{Section}{Sections}
\usepackage{makecell}
\usepackage[normalem]{ulem}

\usepackage{array}
\newcolumntype{L}[1]{>{\raggedright\let\newline\\\arraybackslash\hspace{0pt}}p{#1}}
\newcolumntype{C}[1]{>{\centering\let\newline\\\arraybackslash\hspace{0pt}}p{#1}}
\newcolumntype{R}[1]{>{\raggedleft\let\newline\\\arraybackslash\hspace{0pt}}p{#1}}

\definecolor{pink}{RGB}{255,0,255}

\definecolor{ginger}{RGB}{255,150,0}

\newcommand{\rev}[1]{#1}

\begin{document}

\title{Automated verification of countermeasure against detector-control attack in quantum key distribution}
	
\author{Polina~Acheva}
\email{achevap17@yandex.ru}
\affiliation{Russian Quantum Center, Skolkovo, Moscow 121205, Russia}

\author{Konstantin~Zaitsev}
\affiliation{Russian Quantum Center, Skolkovo, Moscow 121205, Russia}
\affiliation{NTI Center for Quantum Communications, National University of Science and Technology MISiS, Moscow 119049, Russia}

\author{Vladimir~Zavodilenko}
\affiliation{NTI Center for Quantum Communications, National University of Science and Technology MISiS, Moscow 119049, Russia}
\affiliation{HSE University, Moscow 101000, Russia}

\author{Anton~Losev}
\affiliation{NTI Center for Quantum Communications, National University of Science and Technology MISiS, Moscow 119049, Russia}

\author{Anqi~Huang}
\affiliation{\mbox{Institute for Quantum Information \& State Key Laboratory of High Performance Computing, College of Computer Science} \mbox{and Technology, National University of Defense Technology, Changsha 410073, People's Republic of China}}% Country name is spelled this way per APS style

\author{Vadim~Makarov}
\affiliation{Russian Quantum Center, Skolkovo, Moscow 121205, Russia}
\affiliation{NTI Center for Quantum Communications, National University of Science and Technology MISiS, Moscow 119049, Russia}
	
\date{\today}
	
\begin{abstract}
Attacks that control single-photon detectors in quantum key distribution using tailored bright illumination are capable of eavesdropping the secret key. Here we report an automated testbench that checks the detector's vulnerabilities against these attacks. We illustrate its performance by testing a free-running detector that includes a rudimentary countermeasure measuring an average photocurrent. While our testbench automatically finds the detector to be controllable in a continuous-blinding regime, the countermeasure registers photocurrent significantly exceeding that in a quantum regime, thus revealing the attack. We then perform manually a pulsed blinding attack, which controls the detector intermittently. This attack is missed by the countermeasure in a wide range of blinding pulse durations and powers, still allowing to eavesdrop the key. We make recommendations for improvement of both the testbench and countermeasure.
\end{abstract}
	
\maketitle

\section{Introduction}
\label{sec:intro}

In recent years, quantum key distribution (QKD) \cite{bennett1984} has attracted significant attention by both science and industry. This interest is based on its security guaranteed by the laws of quantum mechanics. However, differences exist between the theory of QKD and its practical realisation. They can be exploited by an eavesdropper Eve to steal secret information. For example, a laser pulse attenuated to contain less than one photon on average still sometimes contains two or more photons, which contradicts theoretical assumptions in QKD. A photon-number-splitting attack that exploits these multi-photon pulses was historically the first QKD loophole shown in the year 2000 \cite{brassard2000}. Since then, over twenty different other loopholes have been discovered \cite{makarov2006,gisin2006,qi2007,lamas-linares2007,makarov2008,zhao2008,lydersen2010a,lydersen2010b,li2011a,wiechers2011,lydersen2011c,lydersen2011b,gerhardt2011,sun2011,jain2011,bugge2014,huang2015,sun2015,sajeed2015a,huang2016,sajeed2016,makarov2016,huang2018,zheng2019,huang2020}.

Most loopholes can be closed by additional countermeasures implemented in QKD components or postprocessing. The photon-number-splitting attack \cite{brassard2000} can be closed by a decoy-state protocol \cite{lo2005}; Trojan-horse attack \cite{gisin2006} and laser-seeding attack \cite{sun2015} can be closed by adding isolators to the optical scheme \cite{lucamarini2015,tamaki2016,wang2018a}; detector efficiency mismatch problems \cite{makarov2008,zhao2008,li2011a,huang2015} can be solved by both a proper device calibration and an update to the theory \cite{fung2009}. However, some loopholes have not yet been fully solved, notably the detector control loophole \cite{lydersen2010a,lydersen2010b,wiechers2011,lydersen2011c,lydersen2011b,gerhardt2011,huang2016}. 

The single-photon detector (SPD) currently seems to be the most unsafe QKD system element. There are several attacks focusing on it. Some of them create and exploit mismatch in photon detection efficiency between two or more detectors in the receiver Bob. These include efficiency mismatch in the time domain \cite{makarov2008,zhao2008}, wavelength \cite{li2011a,huang2015}, spatial mode \cite{sajeed2015a}, and during a deadtime \cite{weier2011}. Other attacks control the detector deterministically while blinding it with bright light \cite{lydersen2010a,lydersen2010b,lydersen2011c} or injecting bright pulses at the closing edge of a detector gate \cite{lydersen2011b,qian2018} or in-between the gates \cite{wiechers2011}. The detector control attack was first proposed in 2009 \cite{makarov2009} and found to be applicable to commercial QKD systems the following year \cite{lydersen2010a}. A protection against the attacks on detectors is difficult because Bob has to receive all light from a transmission line with as low loss as possible. (In contrast, a sender Alice can be effectively isolated against attacks that inject light \cite{lucamarini2015,tamaki2016,wang2018a,ponosova2022}.)

Several countermeasures against the bright-light attacks on detectors have been proposed \cite{lydersen2010a,yuan2011,lo2012,silva2012,lim2015,gras2020,sidki2018,jun2015,koehler-sidki2019,alhussein2019,maroy2017,lee2016,qian2019,zhang2021,parra2021}. The most radical one is a measurement-device-independent (MDI) QKD scheme that eliminates the detectors, and thus all their vulnerabilities, from the secure equipment \cite{lo2012}. However, it is less convenient and more costly for commercial implementation than the standard QKD schemes. Other approaches vary in their maturity and effectiveness. \rev{An optical power meter at Bob's entrance with a classical threshold \cite{lydersen2010a} is not fast enough and may overlook a pulsed blinding attack.} A random-detector-efficiency patch \cite{lim2015} was shown to contain unrealistic assumptions on hardware after a careful investigation \cite{huang2016}. A measurement of coincidence click rates \cite{alhussein2019} and application of a random optical attenuation \cite{qian2019} are at a proof-of-principle stage and need further tests. Optical power limiters \cite{zhang2021,parra2021} are not mature and sensitive enough to become a countermeasure. A more mature technology is the measurement of photodiode current to sense the blinding \cite{yuan2011,gras2020}, which is implemented in some commercial SPDs \cite{gras2020,chen2021}. However, its effectiveness as a countermeasure in QKD depends on implementation details and needs to be tested.

For a wide adoption of QKD as a data protection technology, it needs to have certification \cite{langer2009}. The certification standards for QKD include tests for the quality of countermeasures against the known vulnerabilities \cite{iso23837-2022}. \rev{Formalising and} automating the testing procedure would both simplify its application in a certification lab and reduce human factors. We are therefore developing an automated testbench and algorithm that tests SPDs against the bright-light attacks.

The contribution of this paper is two-fold. First, we report an automated setup for testing SPDs against the bright-light attacks. Second, we apply this setup to an SPD with the current-measurement countermeasure. The latter is proven effective against the attack that uses continuous-wave (cw) blinding \cite{gras2020}. This is confirmed with our automated testbench. However, this countermeasure might miss an attack that blinds the detector intermittently by light pulses \cite{gras2020,gao2022}. We probe experimentally in a manual regime the limits of the existing countermeasure implementation. We then make recommendations for both countermeasure and testbench improvement \rev{that would hopefully make them complete and ready for certification}.

\rev{The development of a complete countermeasure is non-trivial. In more than ten years elapsed since the discovery of these attacks \cite{lydersen2010a}, no countermeasure for non-MDI QKD systems has been independently tested and certified as secure. Although it is obvious in the hindsight that the rudimentary countermeasure we test here is insufficient, this is not clear to an engineer designing it without the help of independent testers. We have chosen to focus our development on this type of countermeasure, because it is the simplest and cheapest to implement (being just some extra electronics in the SPD) and it has a potential to close this class of loopholes. However our test methodology may be adopted to other types of countermeasures.}

The paper is organised as follows. In \Cref{sec:setup} we describe the testbench setup, its software, and the detector under test. In \Cref{sec:results} we report experimental results and simulate the attack. We discuss and conclude in \cref{sec:conclusion}.

\section{Experimental setup}
\label{sec:setup}

The SPD control attack using bright light can be realised in several ways. We distinguish three main types of it: continuous blinding \cite{makarov2009,lydersen2010a}, pulsed blinding \cite{lydersen2011c,tanner2014}, and after-gate attack \cite{wiechers2011}. Under the continuous blinding attacks, a cw laser light is applied to the SPD, which is then controlled continuously. Under pulsed blinding, the SPD is blinded and controlled for a period of time longer than the SPD's deadtime. The after-gate attack exploits controllability of gated SPDs in-between the gates, sending short bright pulses outside the gates. We think that testing for all three types of attacks can be automated. Here we demonstrate the automated testing for continuous blinding. We perform the pulsed blinding manually, to better understand the requirements for its automation. The after-gate attack is not applicable to a free-running SPD chosen for our experiment.

\subsection{Automated testbench}
\label{sec:testbench}

\begin{figure}
	\includegraphics{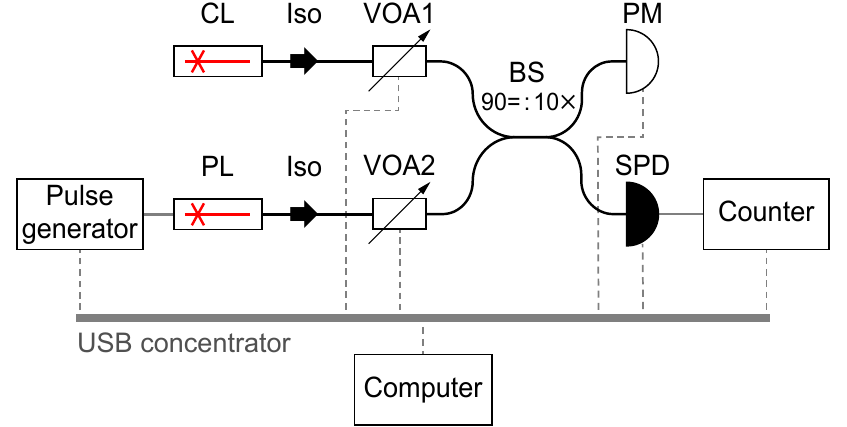}
	\caption{Setup for testing detector control by bright light. CL, continuous-wave laser ($1552~\nano\meter$, $40~\milli\watt$, Thorlabs SFL1550P); PL, pulsed laser ($1552~\nano\meter$, Gooch \& Housego AA1406); Iso, optical isolator; VOA, programmable variable optical attenuator (OZ~Optics DA-100); BS, fiber beamsplitter; PM, optical power meter (Thorlabs PM400 with S155C head); SPD, single-photon detector under test. The pulse generator (Highland Technology P400) drives PL directly and can induce relaxation-limited short laser pulses. The counter (Stanford Research Systems SR620) typically accumulates clicks over $1~\second$ for each data point.}
	\label{fig:blinding-scheme}
\end{figure}

Our testbench setup is shown in \cref{fig:blinding-scheme} \cite{lydersen2010a}. It uses two lasers, a pulsed one and a cw one. Light from each of them passes through an isolator for stability reasons and then a programmable attenuator. Attenuated light from both lasers is then combined on a $90\!:\!10$ beamsplitter, whose outputs are connected to an optical power meter and the detector under test. A computer controls all the devices, runs a testing algorithm, and analyses the data.

\subsection{Software and methodology}
\label{sec:software}

Our software for the automated testbench is written in LabVIEW. It works in two stages. At the first stage, the testbench blinds the SPD by the cw laser. At the second stage, it attempts to control the SPD by the pulsed laser (this method is similar to earlier manual experiments \cite{lydersen2010a}). The program then saves a printable PDF report, an example of which is given in \cref{sec:program-report}, and all the raw data collected. For this particular report example, the entire test sequence took about $1.5~\hour$.

During the first stage, the program uses CL to apply cw power at the SPD, while PL is turned off. The attenuation of VOA1 is scanned through its full $60$ to $0~\deci\bel$ range by a user-settable step ($1~\deci\bel$ in our case). The power is measured by the PM and varies from approximately $2.3\times10^{-11}~\watt$ (near the sensitivity limit of the PM) to $1.25\times10^{-5}~\watt$. At each power level, the detector click rate (measured by the counter) and photocurrent monitor readout value (explained in the next subsection) are recorded. If the click rate drops to zero, the SPD is considered to be blinded.

If the blinding is recorded at one or more power levels, the program proceeds to the second stage. It steps VOA1 again from the maximum attenuation at which the blinding has been recorded through $0~\deci\bel$. At each power level, it also applies short---$240~\pico\second$ full-width at half-magnitude (FWHM)---pulses from the PL at $10~\kilo\hertz$ rate while scanning the attenuation of VOA2 from $60$ to $0~\deci\bel$ by a user-settable step ($1~\deci\bel$ in our case). The energy $E$ of these control pulses is \rev{pre-calibrated} and varies from $10^{-18}$ to $10^{-12}~\joule$. The detector click rate and monitor readout value are again recorded at each energy level. The program then analyses whether the detector clicks with above-zero probability in response to these control pulses (if it does, the SPD is declared controllable in the report) and if a change of $E$ by $3~\deci\bel$ or less leads to the change of click probability from $0$ to $100\%$. The latter is a sufficient condition for a perfect attack on Bennett-Brassard 1984 (BB84) QKD protocol \cite{lydersen2010a}. The report also contains monitor readout plots under control, which may be analysed manually by the operator to see if the countermeasure is effective.

\subsection{Detector under test}
\label{sec:DUT}

\begin{figure}
	\includegraphics{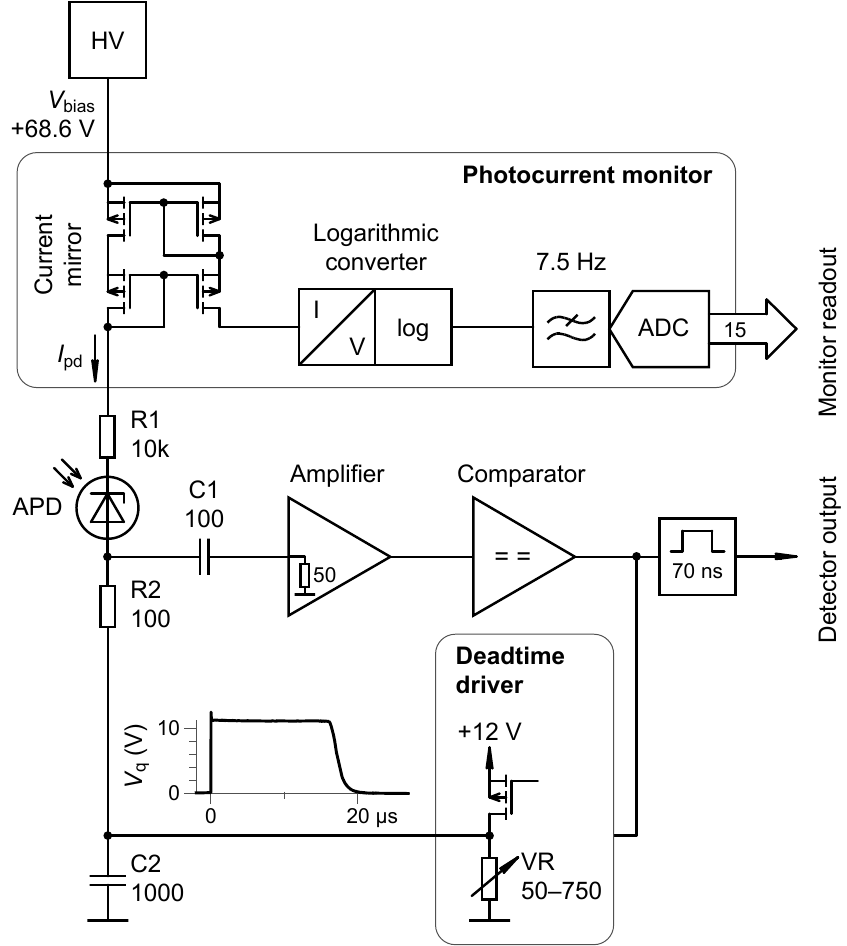}
	\caption{Functional schematic diagram of the single-photon detector. See text for details.}
	\label{fig:functional-circuit}
\end{figure}

In this work, we investigate a free-running single-photon detector manufactured by QRate (serial number \mbox{3-054).} This detector does not use gating, which makes \rev{it} easy to use in a versatile educational kit \cite{rodimin2019} (whereas QRate's commercial QKD system employs a different detector model with sinusoidal gating that has improved performance \cite{losev2022}). Our free-running detector is based on an InGaAs/InP fiber-pigtailed APD (Wooriro WPACPGMOSSNCNP serial number PA19H262-0052) thermoelectrically cooled to $-35~\celsius$. The detector circuit uses passive quenching with enforced deadtime (\cref{fig:functional-circuit}) \cite{stipcevic2017,losev2022,losev2022a}. For Geiger-mode operation, the voltage across the APD should exceed its breakdown voltage by about $2~\volt$. A high-voltage supply (HV; based on Maxim Integrated MAX1932) applies $V_\text{bias} = +68.6~\volt$ at the cathode of the APD via a current mirror and bias resistor R1. When the detector is waiting ready for an avalanche, a stray capacitance between the APD cathode and the circuit ground is charged to the same voltage. This capacitance, on the order of $1~\pico\farad$, is not shown in the circuit diagram but is essential for the detector operation. Once the avalanche begins, the capacitance supplies its current and discharges via the APD and low-impedance circuits connected at the APD anode. The voltage across the APD quickly drops; once it about equals the breakdown voltage, the current reduces to a value when the avalanche no longer self-sustains and the avalanche then stops \cite{kim2011}. Note that in this passively-quenching circuit, the current supplied from HV via R1 is not sufficient to sustain the avalanche. It merely recharges the stray capacitance relatively slowly to $V_\text{bias}$ after the avalanche quenches.

The onset of the avalanche current is sensed by an amplifier (Analog Devices HMC589AST89E) and high-speed comparator (Analog Devices ADCMP573), then expanded in duration to $70~\nano\second$ with a single-shot generator, producing a logic signal at the output of the detector. To reduce afterpulsing, an enforced deadtime $\tau = 20~\micro\second$ is applied by raising the voltage at the APD anode by $11.3~\volt$. (This also ensures ending any occasional avalanche that does not cease via the passive quenching \cite{kim2011}.) The deadtime driver removes this voltage at the end of the deadtime gradually via a variable resistor VR (implemented with a series of transistor switches), to avoid triggering additional avalanches \cite{losev2022,losev2022a}. Once this process is complete, the stray capacitance charges via R1 to $V_\text{bias}$ and the detector becomes ready for the next avalanche. The photon detection efficiency of our detector sample is $2.2\%$ at $1550~\nano\meter$ and the dark count rate $D = 412~\hertz$.

\rev{We remark that a commercial QKD system would use a gated detector with a higher photon detection efficiency, such as the sinusoidally-gated detector \cite{losev2022}, which we have also tested on this testbench without a countermeasure \cite{makarov2023}. However this free-running detector sample is sufficient for the initial development of our testbench and the countermeasure. Its low photon detection efficiency does not affect the test methodology. In the future, the countermeasure will also be implemented and tested in QRate's sinusoidally-gated detector.}

\begin{figure}
	\includegraphics{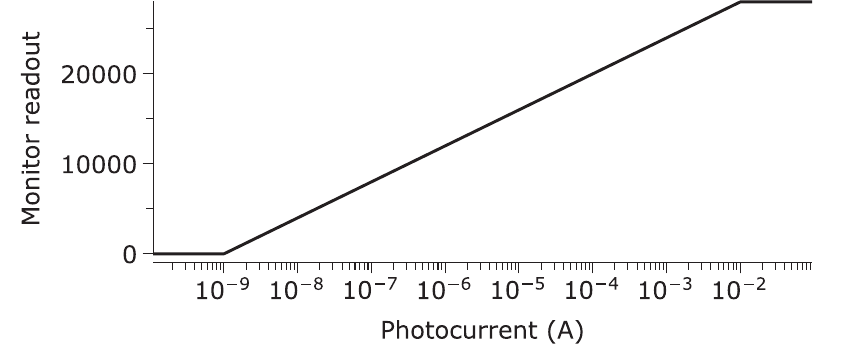}
	\caption{Monitor readout $M$ as a function of a constant APD current $I_\text{pd}$. The curve has been theoretically calculated based on the data sheets of the integrated circuits.  Note that with $V_\text{bias} = +68.6~\volt$ and the resistor values used in this particular detector sample, $I_\text{pd}$ cannot exceed about $6.8~\milli\ampere$.}
	\label{fig:detector-countermeasure-vs-current}
\end{figure}

To provide the countermeasure against detector blinding \cite{lydersen2010a}, \rev{the present free-running} SPD employs a photocurrent monitor circuit (\cref{fig:functional-circuit}) \cite{yuan2011,gras2020}. The current $I_\text{pd}$ flowing into R1 is copied by a current mirror, processed by a logarithmic converter (Analog Devices AD8305) configured with a reduced bandwidth, and digitised by an analog-to-digital converter (ADC; Microchip MCP3425) with 15-bit resolution. The analog-to-digital converter outputs a readout value
\begin{equation}
\label{eq:monitor-readout}
M = 4000\log_{10}\left(\frac{I_\text{pd}}{1~\nano\ampere}\right),
\end{equation}
which obeys this equation in a wide range of constant-current values from $10^{-9}$ to $10^{-2}~\ampere$ (\cref{fig:detector-countermeasure-vs-current}). The logarithmic signal passes a low-pass frequency filter with $7.5~\hertz$ rolloff (intrinsic to the ADCs of delta-sigma type) and is digitised at $15~\hertz$ rate. The latest readout value is made available via a universal serial bus (USB) interface to a computer running either a detector monitoring software supplied by the manufacturer or user-written program like our testbench automation. The latter records a single sampled value whenever it takes a data point for the automatically generated report. These single values of $M$ fluctuate significantly when the detector is producing random counts (the fluctuation is up to $\pm 700$ for dark counts, less at higher count rates). To reduce these random fluctuations, in manual measurements of pulsed blinding we have done additional averaging for each data point, sampling 30 readout values spread evenly over $14.5~\second$ and calculating their mean.

Note that the $7.5~\hertz$ low-pass filter in the monitor circuit is placed \emph{after} the logarithmic converter, which itself outputs a signal with a much higher bandwidth of the order of $1~\kilo\hertz$. When $I_\text{pd}$ varies in time at a frequency faster than $7.5~\hertz$ but below $1~\kilo\hertz$, this circuit does not average the photocurrent but rather averages its logarithm. The readout $M$ then \emph{underestimates} the mean value of $I_\text{pd}$. This helps Eve in defeating this countermeasure, as we show below.

\section{Experimental results}
\label{sec:results}

\subsection{Countermeasure calibration}
\label{sec:countermeasure-calibration}

\begin{figure}
	\includegraphics{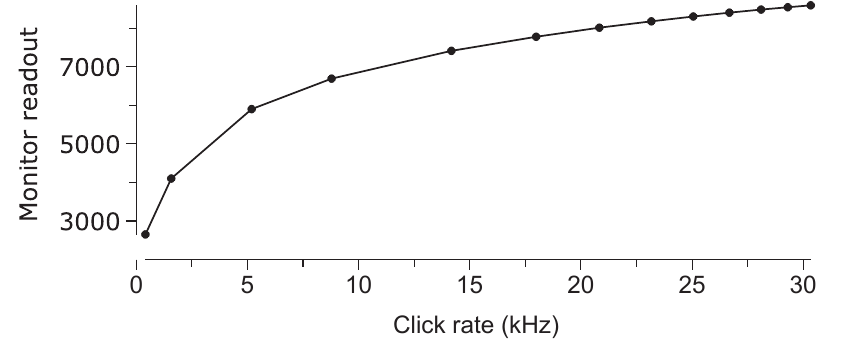}
	\caption{Monitor readout $M$ under detector illumination by $0.8$-photon pulses. The leftmost point is for unilluminated detector with dark counts only, while the other points are under illumination at a laser pulse rate from $10~\kilo\hertz$ to $1~\mega\hertz$. An estimated ``normal'' detector click rate of $20~\kilo\hertz$ or less results in single sampled values of $M < 8100$.}
	\label{fig:detector-countermeasure}
\end{figure}

Before the monitor readout can be used to detect attacks, we need to estimate the values of $M$ observed during normal operation of the QKD system. We simulate the single-photon regime by illuminating the SPD with laser pulses attenuated to $0.8$~photon/pulse at a varying rate. The results are shown in \cref{fig:detector-countermeasure}. We assume a typical detector click rate in a modern QKD system is $20~\kilo\hertz$ \footnote{In QRate QKD proof-of-principle experiments \cite{duplinskiy2017}, the system runs at $10~\mega\hertz$ source pulse rate over two lines: a $50~\kilo\meter$ fiber spool and $30~\kilo\meter$ urban line. Scaling the click rates reported to a future $1~\giga\hertz$ source rate, we expect a single detector count rate of $50$ and $13~\kilo\hertz$. We thus assume $20~\kilo\hertz$ to be a typical click rate.}. The monitor readout is then expected not to exceed $8100$ for single sampled values (or $7900$ averaged). Any value larger than that indicates an attack.

\subsection{Continuous-wave blinding of detector}
\label{sec:results-blinding-cw}

\begin{figure*}
	\includegraphics{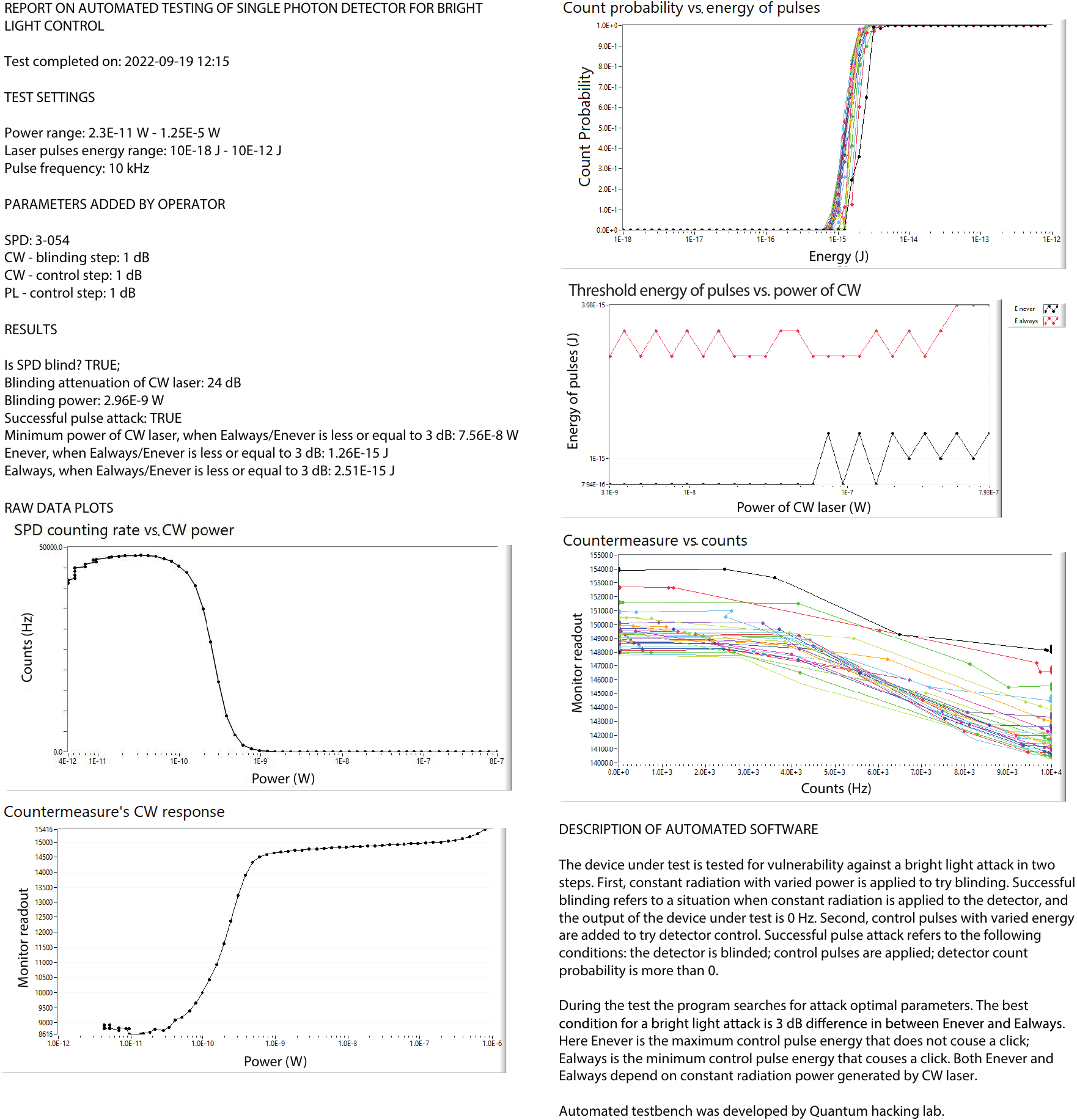}
	\caption{Report generated by software after automated test of cw blinding and control. \rev{The multiple curves in the third and fifth plots are taken at cw power values shown in the fourth plot.} See text for details.}
	\label{sec:program-report}
\end{figure*}

The continuous blinding attack is executed automatically by the testbench. The report (\cref{sec:program-report}) includes the following plots: the dependence of the count rate and monitor readout value $M$ on cw laser power (with PL off), the probability that the blinded SPD produces a click versus control pulse's energy $E$ (each curve is at a different cw blinding power), the maximum pulse energy that never produces a click $E_\text{never}$ and minimum energy that always produces a click $E_\text{always}$ \cite{lydersen2010a} versus the cw blinding power, and the dependence of $M$ on the SPD click rate when it is being blinded and controlled (each curve is at a different cw blinding power). The software automatically analyses the click rates and makes conclusions that this detector is blindable and the click probability under control is non-zero. \rev{Note that the cw power and $E$ are both scanned an order of magnitude or more above and below the values where the full control is observed.}

We can manually analyse the monitor readout plots and see that the countermeasure catches this attack. When the SPD is blinded at $2.96~\nano\watt$, the countermeasure registers $M \approx 14800$, and under total control at $75.6~\nano\watt$ $M \approx 15000$. This significantly violates the safety condition $M \leq 8100$ calibrated in \cref{sec:countermeasure-calibration}.

\begin{figure}
	\includegraphics{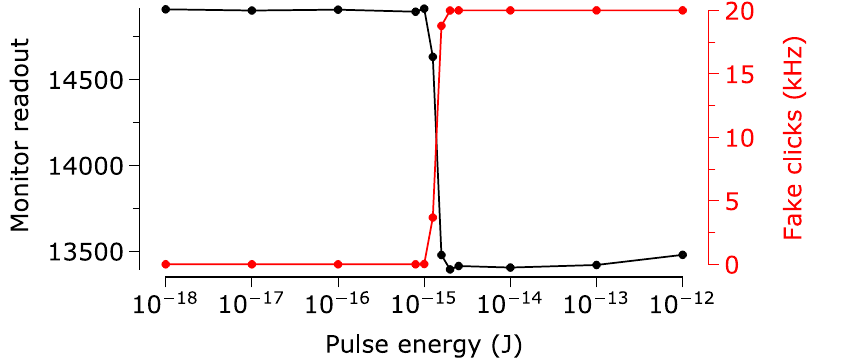}
	\caption{Monitor readout and click rate versus trigger pulse energy $E$, at $31.5~\nano\watt$ cw blinding power and $20~\kilo\hertz$ trigger pulse rate. The countermeasure is not affected by $E$ until it causes a click.}
	\label{fig:counter_vs_energy}
\end{figure}

For completeness, we also need to check $M$ when the blinded SPD is controlled by the PL. The monitor readout drops perceptibly when the detector begins to click, while being otherwise independent on $E$ (see \cref{fig:counter_vs_energy} and the last plot in \cref{sec:program-report}). The drop of $M$ can be explained by the forced reduction of voltage across the APD during the deadtime, which decreases the APD's internal gain and thus its mean photocurrent in response to cw illumination. However, even the reduced $M \geq 13405$ remains well above the ``normal'' monitor readout of $8100$. This keeps the countermeasure effective against the continuous blinding.

\rev{We observe no temporary or permanent deterioration of the SPD during our tests. This type of APD-based detector is known to withstand $10~\milli\watt$ cw optical illumination without damage \cite{lydersen2010b}, which is higher than the optical power in our testbench.}

\subsection{Pulsed blinding of detector}
\label{sec:results-blinding-pulsed}

\begin{figure}
	\includegraphics{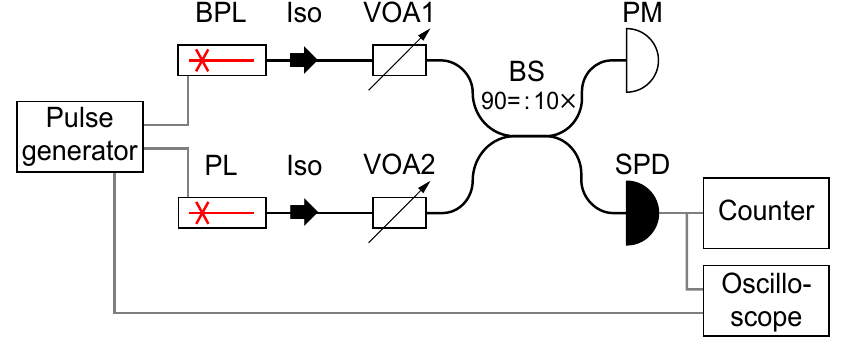}
	\caption{A modified setup for testing detector control under pulsed blinding. BPL, blinding pulse laser [$1552~\nano\meter$, $40~\milli\watt$, Allwave Lasers SWLD-1550-100-PM(DBF)]. The pulse generator drives BPL directly and induces long laser pulses. The pulsed laser (PL) is also driven by the pulse generator directly and emits a relaxation-limited short laser pulse with $240~\pico\second$ FWHM delayed in respect to the start of the BPL pulse. The counter typically accumulates clicks over $1~\second$ for each data point.}
	\label{fig:pulse-blinding-scheme}
\end{figure}

\begin{figure}
	\includegraphics{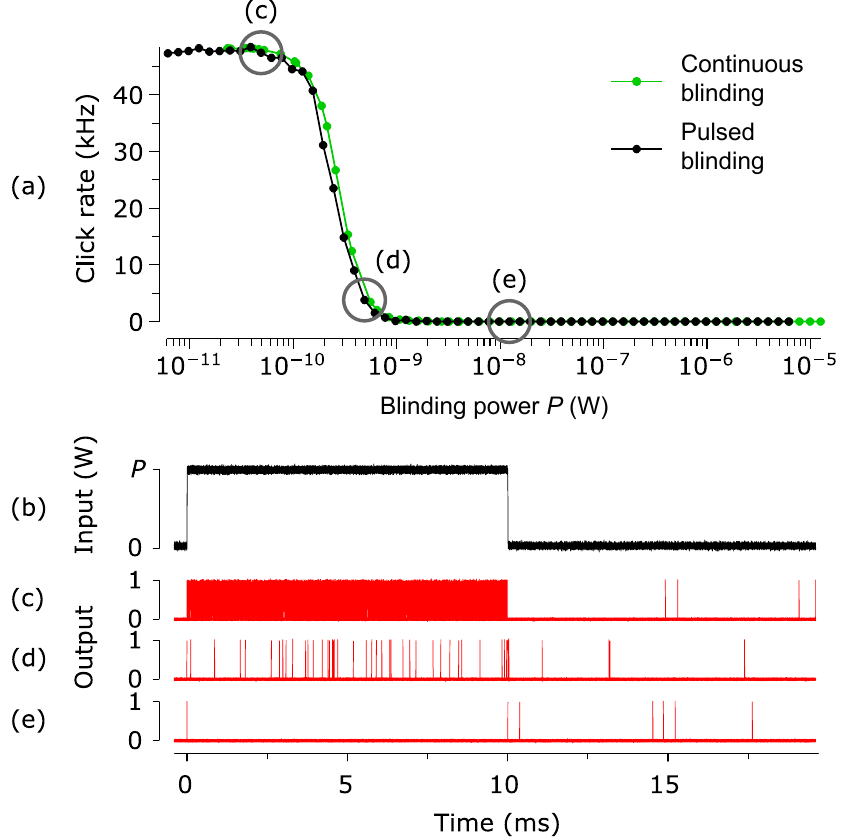}
	\caption{Pulsed blinding of the SPD. (a)~Click rate under blinding. For pulsed blinding, only the click rate within the blinding pulse is measured; outside the blinding pulse, the detector works in the normal photon counting mode with dark counts. Click rate under cw blinding is shown for comparison. (b)~The blinding pulse of peak power $P$. (c)~Oscillogram of detector output ($70~\nano\second$ long logic pulses of $1.5~\volt$ amplitude) when $P = 49~\pico\watt$ is insufficient for blinding and causes a saturated click rate within the pulse. (d)~At a higher $P = 490~\pico\watt$, the click rate within the pulse drops significantly. (e)~Detector is blinded within the pulse of $P = 12.3~\nano\watt$.}
	\label{fig:pulse-blinding-graph}
\end{figure}

The photocurrent-measuring countermeasure may be ineffective against the pulsed blinding attack \cite{gras2020,gao2022}. In this attack, the SPD is blinded temporarily and controlled while blinded. It works in the normal photon counting mode between the blinding pulses. We modify our experimental setup slightly (\cref{fig:pulse-blinding-scheme}) and use it to manually test the SPD. We first apply $10~\milli\second$ long blinding pulse of peak power $P$ \rev{(measured by PM)} at $20~\hertz$ repetition rate and observe the detector clicks. The results are shown in \cref{fig:pulse-blinding-graph}. We use an oscilloscope ($3.5~\giga\hertz$ bandwidth; LeCroy 735Zi) to select the clicks that occur during the blinding pulse and measure their rate. The complete blinding within the pulse occurs at $P = 2.46~\nano\watt$, which is almost the same power as in the cw blinding ($2.96~\nano\watt$).

\begin{figure}
	\includegraphics{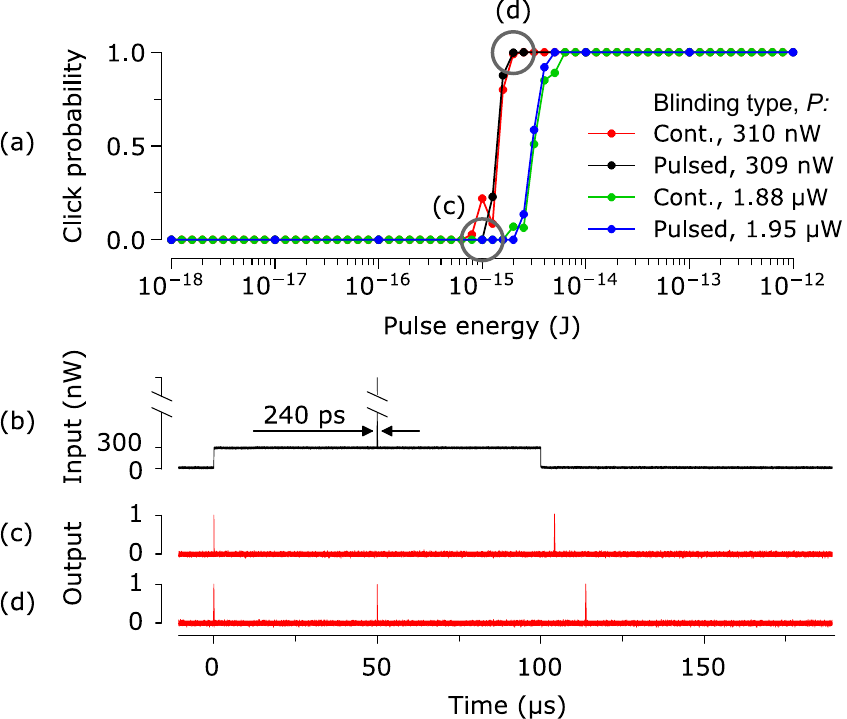}
	\caption{Control of SPD within $100~\micro\second$ long blinding pulses applied at $20~\hertz$ rate. (a)~Click probability versus trigger pulse energy $E$. Click probabilities under cw blinding are shown for comparison. The probabilities are measured over $10^{3}$ pulses. (b)~The blinding and trigger pulses, the latter being of $240~\pico\second$ FWHM and applied $50~\micro\second$ after the start of the blinding pulse. (c)~Detector output at $P = 309~\nano\watt$ and $E = 10^{-15}~\joule$. This trigger pulse never causes a click. (d)~The trigger pulse energy is increased by $3~\deci\bel$ to $2\times10^{-15}~\joule$. It always causes a click.}
	\label{fig:click_probability}
\end{figure}

\begin{figure}
	\includegraphics{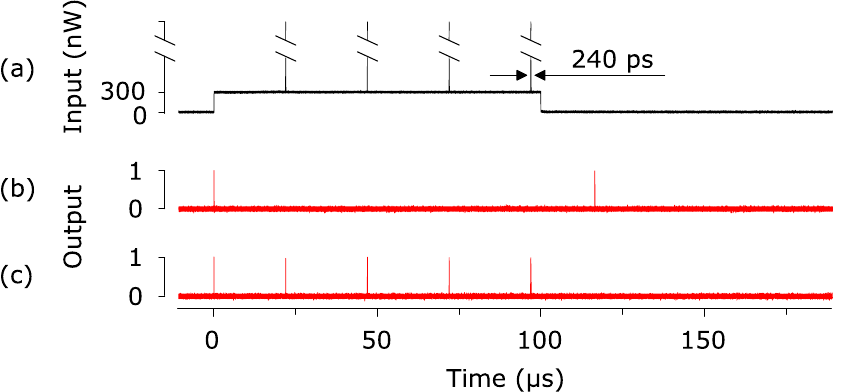}
	\caption{Detector control by multiple trigger pulses applied during one blinding pulse. (a)~Blinding and trigger pulses, the latter being of $240~\pico\second$ FWHM. The blinding pulse has $P = 309~\nano\watt$ and is applied at $20~\hertz$ rate. (b)~Detector output at $E = 10^{-15}~\joule$. These trigger pulses never cause clicks. (c)~The trigger pulses' energy is increased by $3~\deci\bel$ to $2\times10^{-15}~\joule$. Each of them always causes a click.}
	\label{fig:fake-pulses}
\end{figure}

The detector behaviour within the pulse closely resembles that under the cw blinding. If an additional short trigger pulse of energy $E$ is applied during the blinding pulse, it responds with a click (\cref{fig:click_probability}). Less than $3~\deci\bel$ change of $E$ is required to transition between $0$ and $100\%$ click probability. Note that the detector always clicks at the start of the blinding pulse, and often also at the end of it. Because of these additional uncontrollable clicks, it is beneficial for Eve to apply multiple trigger pulses during the blinding pulse. In \Cref{fig:fake-pulses}, control by four trigger pulses is illustrated. The response to the trigger pulse does not depend on its timing within the blinding pulse; the click probability changes less than $2\%$ throughout. We have also verified that up to 199 trigger pulses spaced $20~\micro\second$ apart (i.e.,\ the exact length of the detector deadtime $\tau$) applied during a longer $4~\milli\second$ blinding pulse work as well. The monitor readout decreases slightly as the number of triggered clicks increases, similarly to the effect observed in \cref{fig:counter_vs_energy}. 

\begin{figure}
	\includegraphics{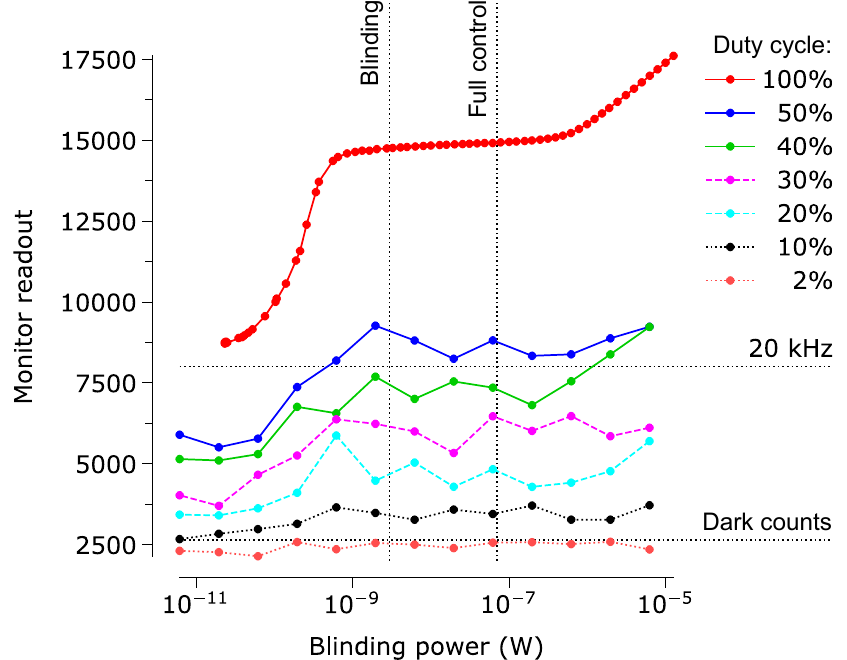}
	\caption{Monitor readout under pulsed blinding at different duty cycle values of the blinding illumination, at $20~\hertz$ repetition rate of the blinding pulses. $100\%$ is cw blinding, $50\%$ is $25~\milli\second$ long blinding pulse, $40\%$ is $20~\milli\second$ long blinding pulse, etc. Dotted horizontal lines indicate an expected countermeasure readout in the normal photon counting mode at $20~\kilo\hertz$ click rate and at the dark count rate (as calibrated in \cref{sec:countermeasure-calibration}). Dotted vertical lines indicate the minimum blinding power and minimum cw power at which $E_\text{always}/E_\text{never} \leq 2$.}
	\label{fig:countermeasures}
\end{figure}

Finally, we check how the countermeasure responds to the pulsed blinding of different duty cycle values. We vary the blinding pulse width while keeping its repetition rate constant at $20~\hertz$, see \cref{fig:countermeasures}. At this repetition rate, the blinding pulse can be as long as $20~\milli\second$ without causing an abnormally high monitor readout of more than $7900$. The low monitor readout under the pulsed blinding is partially explained by the unwisely constructed sequence of first taking the logarithm then averaging at the low-pass frequency filter in the photocurrent monitor circuit (\cref{sec:DUT}). This implementation of the countermeasure is thus unable to detect the pulsed blinding of up to $40\%$ duty cycle, which leaves Eve ample room for attack.

\subsection{Intercept-resend attack model}
\label{sec:attack-model}

The experimental results on pulsed blinding show that Eve has a significant degree of control over the SPD, while not being revealed by the countermeasure. This SPD behaviour violates the assumptions on a measurement apparatus made in most security proofs for QKD, in particular the independence of detection probability on Bob's basis choice \cite{lydersen2011b}, rendering these proofs inapplicable. We thus clearly cannot guarantee the security of QKD that employs such SPDs, regardless of whether we know how to construct Eve's attack in detail or not.

Nevertheless, here we attempt to model such attack. We assume the QKD system runs the BB84 protocol \cite{bennett1984} with an active basis choice and two detectors at Bob. Eve intercepts Alice's output at the beginning of the lossy quantum channel using a receiver with a high detection efficiency and very low error rate. She then resends blinding pulses and faked states to Bob according to her measurement results \cite{lydersen2010a}. We also assume Bob's dark counts are the only source of errors in the system without Eve. \rev{Let us} approximately estimate Alice's and Bob's quantum bit error rate (QBER) under attack and the rate at which Eve can trigger clicks at Bob.

We consider one pulsed blinding period of length $T$, consisting of $CT$ blinding time and $(1-C)T$ idle time, where $C \in (0, 1)$ is the duty cycle. The blinding pulse causes a simultaneous click in both Bob's detectors at its start and, possibly, a click at its end. We assume these, on average, record in the raw key as two clicks, of which one is erroneous. During the blinding time Eve can induce $\gtrsim CT/2\tau$ controlled clicks at Bob (the exact number depends on Eve's detection rate and whether she can send faked states during Bob's deadtime). Bob also registers about $2(1-C)TD$ dark counts during the idle time. Bob's click rate
\begin{equation}
\label{eq:intercept-resend-rate-Bob}
R_\text{B} \approx \frac{2 + CT/2\tau + 2(1-C)TD}{T}
\end{equation}
and
\begin{equation}
\label{eq:intercept-resend-QBER}
\text{QBER} \approx \frac{1+(1-C)TD}{TR_\text{B}}.
\end{equation}
We stress that the above calculation is approximate and ignores lesser effects like double clicks at Bob, rate reduction owing to his detector deadtime, sources of errors other than Bob's dark counts, etc.

Taking the experimental parameters from this paper ($T = 50~\milli\second$, $C = 0.4$, etc.),\ we get $R_\text{B} \approx 10.5~\kilo\hertz$ and $\text{QBER} \approx 2.5\%$. These are reasonable parameters for a QKD system and it should generate a key. Since Eve is performing the intercept-resend attack, the generation of secret key under this attack is in fact impossible \cite{curty2004a}.

\rev{The main limitation of this attack is its ability to replicate $R_\text{B}$ expected by the legitimate users. Its value depends on the system implementation and the line loss, and for many practical settings is less than $10.5~\kilo\hertz$ \cite{duplinskiy2017}. Also a faster QKD system would use detectors with shorter $\tau$, which helps Eve obtain a higher $R_\text{B}$ [\cref{eq:intercept-resend-rate-Bob}].} If Bob's click rate under the attack is \rev{still} insufficient to replicate the system performance expected by Alice and Bob, Eve can choose to bypass a fraction of Alice's photons into the quantum channel to Bob during the idle time. This strategy would not be an optimal attack. We would then need to consider Eve's key information versus the amount of privacy amplification Alice and Bob apply. We speculate that either attack strategy should also work with the decoy-state protocol \cite{lo2005}, as this attack does not drastically affect the yield of different photon-number states.

\section{Discussion and conclusion}
\label{sec:conclusion}

Our testbench has tested the free-running SPD for cw blinding and control and made conclusions about it fully automatically, essentially replicating the well-known manual testing method \cite{lydersen2010a,gras2020}. A manual analysis of collected data shows that the countermeasure reveals the cw blinding reliably. We then manually demonstrate pulsed blinding and control of this SPD. The countermeasure fails to reveal the pulsed blinding of up to $40\%$ duty cycle, allowing Eve to control the detector during the blinding pulses. Our modeling shows that the intercept-resend attack on QKD should then still be possible.

To build the testbench good enough for certification purposes, its automatic operation should be extended to pulsed blinding regimes. The testbench should also automatically analyse the countermeasure output under both cw and pulsed blinding, and make a pass/fail conclusion whether the countermeasure reveals all the attacks. Such extension of the testing algorithm is a topic for future work. In order to develop it, we need to have the SPD with a properly implemented countermeasure that reveals the pulsed attacks.

The existing countermeasure implementation fails to reveal the pulsed blinding primarily because of very low ADC sampling rate of the photocurrent (15~\hertz). Our results suggest that increasing the bandwidth and processing the monitor signal for peak detection would be a step in the right direction. Direct measurements of the signal at the output of the logarithmic converter with an oscilloscope suggest that an ADC with $\sim 1~\mega\hertz$ sampling rate or an analog comparator (i.e.,\ a voltage threshold detector) would be sufficient to reveal the pulsed blinding. The necessary hardware can easily be added to the next version of QRate's free-running SPD. Implementing and testing this improved countermeasure, \rev{as well as adopting it for the sinusoidally-gated detector,} will be our next study. \rev{Testing superconducting-nanowire single-photon detectors with a built-in countermeasure \cite{tanner2014} is also a promising application.}

The quantum key distribution protocol needs to be amended to take input from the countermeasure. One obviously secure method is to discard the entire accumulated raw key and start a new QKD session whenever an abnormally high monitor readout value occurs. A~less wasteful approach might be to discard potentially compromised raw key data in a limited time range that surrounds the abnormally high monitor readout, while continuing the current QKD session. We remark that the countermeasure might occasionally be triggered by benign transient events like electromagnetic interference, computer glitch, or optical line maintenance \cite{huang2016}. If the problem persists over multiple key distillation sessions, it might be a good idea to alert the human operator of the system of this abnormality, which may be caused by a technical malfunction or the actual attempt of attack.

\rev{We finally remark that our testbench does not test for effects that may appear at higher optical power, such as thermal blinding \cite{lydersen2010b} and laser damage of APD \cite{bugge2014}. While the thermal blinding can be tested in this setup, the laser damage requires significant modifications of the testbench \cite{huang2020,ponosova2022}.}

\acknowledgments
We thank Hao Qin for discussions and for providing motivation for this study.

\section*{List of abbreviations}
\label{sec:abbreviations}

\noindent QKD, quantum key distribution;

\noindent SPD, single-photon detector;

\noindent MDI QKD, measurement-device-independent QKD;

\noindent CL, continuous-wave laser;

\noindent PL, pulsed laser;

\noindent Iso, optical isolator;

\noindent VOA, programmable variable optical attenuator;

\noindent BS, fiber beamsplitter;

\noindent PM, optical power meter.

\section*{Declarations}
\label{sec:declarations}

\noindent \textbf{Funding:} This work was funded by the Ministry of Science and Education of Russia (program NTI center for quantum communications and grant 075-11-2021-078), Russian Science Foundation (grant 21-42-00040), the National Natural Science Foundation of China (grants 61901483 and 62061136011), the National Key Research and Development Program of China (grant 2019QY0702), and the Research Fund Program of State Key Laboratory of High Performance Computing (grant 202001-02).

\medskip

\noindent \textbf{Authors' contributions:} P.A.\ and K.Z.\ designed and programmed the automated testbench, performed the experiments, and analysed the data. A.H.\ and V.M.\ analysed the data. V.Z.\ and A.L.\ developed the detector under test and the countermeasure. P.A.,\ K.Z.,\ and V.M.\ wrote the paper with input from all authors. V.M.\ supervised the detector testing project.

\medskip

\noindent \textbf{Ethical approval and consent to participate:} Not applicable.

\medskip

\noindent \textbf{Consent for publication:} All authors have approved the publication. The research in this work did not involve any human, animal or other participants.

\medskip

\noindent \textbf{Availability of supporting data:} Partial data generated or analysed during this study are included in this published article. Any datasets used and/or analysed during the current study that have not been included in this published article are available from the corresponding author on reasonable request.

\medskip

\noindent \textbf{Competing interests:} The authors declare no competing interests.

%\def\bibsection{\medskip\begin{center}\rule{0.5\columnwidth}{.8pt}\end{center}\medskip} % Redefines bibliography separator to single-column. This reduces chances of float placement bugs in the last page.
%\bibliography{library}

\end{document}